\documentclass[12pt]{iopart}
\usepackage{amsmath}
\usepackage{iopams}

\begin{document}

\maketitle

\title{A New Fundamental Asymmetric Wave Equation 
and its Application to Acoustic Wave Propagation}

\author{Z. E. Musielak}
\address{Department of Physics, The University of Texas at 
Arlington, Arlington, TX 76019, USA}
%

\begin{abstract}
The irreducible representations of the extended Galilean group 
are used to derive the symmetric and asymmetric wave equations.  
It is shown that among these equations only a new asymmetric 
wave equation is fundamental.  By being fundamental the equation 
gives the most complete description of propagating waves as it 
accounts for the Doppler effect, forward and backward waves, 
and makes the wave speed to be the same in all inertial frames.  
To demonstrate these properties, the equation is applied to 
acoustic waves propagation in an isothermal atmosphere, and
to determine Lamb's cutoff frequency.  
\end{abstract}


\section{Introduction} 
   
In modern physics, a dynamical equation is called fundamental if  
it is local, has its Lagrangian and remains invariant with respect  
to the spatio-temporal transformations that form a group of the 
metric, and internal symmetries that form a gauge group [1]; the 
latter is related to interactions, which do not affect the description 
of free particles and waves.  The law of inertia for classical 
particles [2-4], and the Schr\"odinger [5] and  L\'evy-Leblond [6] 
equations for quantum particles are examples of nonrelativistic 
fundamental equations.  Moreover, all basic equations of 
relativistic classical and quantum physics are fundamental [1]. 

In nonrelativistic physics, space and time are Galilean and their 
metrics are $ds_1^2\ =\ dx^2 + dy^2 + dz^2$ and $ds_2^2\ 
=\ dt^2$, respectively, with $x$, $y$ and $z$ being the spatial 
coordinates and $t$ being time.  All transformations that leave 
the Galilean metrics unchanged form the Galilean group of the 
metric [6], meaning that the metrics preserve their forms in all 
inertial frames, and observers associated with these frames are 
called {\it Galilean observers}.  

For free particles in Classical Mechanics (CM), the law of inertia 
is fundamental because it is local, has its well-known Lagrangian 
[2-4], and it is also invariant with respect to the Galilean group 
of the metric [6]; this means that for all Galilean observers 
the form of the equation describing this law is the same.  On 
the other hand, the second law of dynamics may or may not 
be Galilean invariant depending on the form of its force [2-4].

The wave equation of classical physics describes the 
propagation of waves in a given background medium [7-9]; 
however, the wave equation is not fundamental because it 
is not Galilean invariant [6,10]. There are two main reasons:
(i) the Galilean metrics require wave equations to be 
asymmetric in space and time derivatives, and (ii) Galilean 
invariance requires the wave speed to be the same in all 
inertial frames, which is violated by any classical wave 
propagating slower than the speed of light.  Both reasons
apply to the wave equation, therefore, it is not fundamental.

The Galilean group of the metric can be extended to make 
its structure similar to that of the Poincar\'e group [11].  Let 
$\mathcal {G}_e$ be the extended Galilean group [10-12] 
with its mathematical structure $\mathcal {G}_e = [O(3) 
\otimes_s B(3)] \otimes_s [T(3+1) \otimes U(1)]$, where 
$O(3)$ and $B(3)$ are subgroups of rotations and boosts, 
respectively.  In addition, $T(3+1)$ is an invariant Abelian 
subgroup of combined translations in space and time, and 
$U(1)$ is a one-parameter unitary subgroup.  The subgroup 
$T(3+1)$ plays an important role in $\mathcal {G}_e$ because 
its irreducible representations (irreps) are well-known [13,14] and 
they provide labels for all the irreps of $\mathcal {G}_e$ [13-15]. 

The Schr\"odinger equation of Quantum Mechanics (QM) is 
Galilean invariant since its form remains the same when all 
transformations of $\mathcal {G}_e$ are applied to it; 
however, the invariance requires that the phase factor is 
introduced [5,16-18].  By being Galilean invariant, local 
and with its Lagrangian known, the Schr\"odinger equation 
is the fundamental equation of QM as its form remains the 
same for all Galilean observers.  Moreover, the scalar 
wavefunction of the equation transforms as one of the
irreps of $\mathcal {G}_e$, which guarantees that all
Galilean observers identify the same physical object 
respresented by the function.  

The  L\'evy-Leblond equation [6] whose spinor wavefunction 
describes elementary particles with spin in nonrelativistic QM 
is linear, has its Lagrangian, and is also Galilean invariant,
which means that it is fundamental.  Thus, the  L\'evy-Leblond 
and Schr\"odinger equations are two fundamental wave 
equations of nonrelativistic QM, and they describe particles 
with and without spin, respectively.

In relativistic classical physics, the wave equation for 
electromagnetic waves is fundamental since the speed
of light remains the same in all inertial frames. However,
there is no similar equation in nonrelativistic physics for
classical waves propagating with speeds lower than the 
speed of light.  Previous attempts to use different forms 
of the Schr\"odinger equation to describe propagation 
of classical waves were made [19-23] but the resulting 
equations were not fundamental.  Therefore, the main 
aim of this paper is to derive such an equation by 
following the recent work [24] in which a new 
asymmetric wave equation was discovered, and used 
to formulate a theory of cold dark matter [25].

In this paper, the conditions for the new asymmetric 
wave equation to become a fundamental wave equation 
for classical waves are established and discussed.  To 
compare the wave description given by the non-fundamental 
and fundamental wave equations, both formulations are 
used to describe the propagation of acoustic waves in an 
isothermal atmosphere and to determine Lamb's cutoff 
frequency.

The paper is organized as follows: in Section 2, the basic 
equations are derived and discussed; the wave equations 
and their Lagrangians are obtained in Section 3; Galilean 
invariance of the wave equations is investigated in Section 
4; applications of the obtained results to acoustic wave 
propagation are presented in Section 5; and conclusions 
are given in Section 6.

\section{Derivation of symmetric and asymmetric equations}

The invariant Abelian subgroup $T(3+1)$ of combined translations 
in space and time plays an important role in $\mathcal {G}_e$ 
because its irreps are well-known [12-15] and they provide labels 
for all the irreps of $\mathcal {G}_e$ [13,14].  The conditions that 
the scalar wavefunction $\phi (t, \mathbf {x})$ transforms as one 
of the irreps of $\mathcal {G}_e$ are given by the following 
eigenvalue equations [17,18] (see the Appendix A for their 
derivation): 
\begin{equation}
i {{\partial} \over {\partial t}} \phi (t, \mathbf {x}) = \omega\ 
\phi (t, \mathbf {x})\ ,
\label{eq1a}
\end{equation}  
and 
\begin{equation}
- i \nabla \phi (t, \mathbf {x}) = \mathbf {k}\ \phi (t, \mathbf {x})\ ,
\label{eq1b}
\end{equation}  
where $\phi (t, \mathbf {x})$ is an eigenfunction of the generators of 
$T(3+1)$, and the eigenvalues $\omega$ and $\mathbf {k}$ are real
constants that label the irreps.  The generator of translation in time is 
$\hat E = i \partial / \partial t$, and the generator of translations in 
space is $\hat P = - i \nabla$, with $[\hat E, \hat P] = 0$ being 
the commuting operators.  The group $\mathcal {G}_e$ also has the 
generator of boosts given by $\hat V = t \hat P$, which means that 
the eigenvalues for the operators $\hat V$ and $\hat P$ must be the 
same [17,18].  The fact that $\phi (t, \mathbf {x})$ obeys Eqs 
(\ref{eq1a}) and (\ref{eq1b}) and transforms as one of the 
irreps of $\mathcal {G}_e$ means that all Galilean observers 
identify the same object, which is a wave under consideration, 
and their description of this wave is identical.                                                                                                                                                                                                    

The obtained eigenvalue equations can be used to derive all wave equations 
of physics for scalar wavefunctions that are allowed to exist in the Galilean 
space and time.  In general, the derived dynamical equations can be divided 
into two separate families, namely, the symmetric equations, with the 
same order of space and time derivatives, and the asymmetric equations, 
with different orders of space and time derivatives [17].  Moreover, the 
equations can be of any order [18], but in this paper only the second-order 
equations are considered.

The only second-order symmetric equation that can be derived from 
the eigenvalue equations is 
\begin{equation}
\left [ {{\partial^2} \over {\partial t^2}} -  C_{1} \nabla^{2} \right ] 
\phi (t, \mathbf {x}) = 0\ , 
\label{eq1}
\end{equation}  
where $C_{1} = \omega^2 / k^{2}$, with $k^{2} = (\mathbf {k} 
\cdot \mathbf {k})$.  Since $C_{1}$ is a real constant coefficient 
of arbitrary value, there is an infinite set of these second-order 
equations, and they are called {\it wave-like equations} [24]. 

Two different asymmetric second-order equations resulting from Eqs.
(\ref{eq1a}) and (\ref{eq1b}) can also be obtained [24]:   
\begin{equation}
\left [ i {{\partial} \over {\partial t}} + C_{2} \nabla^{2} \right ] 
\phi (t, \mathbf {x}) = 0\ ,
\label{eq2}
\end{equation}  
and 
\begin{equation}
\left [ {{\partial^2} \over {\partial t^2}} -  i  C_{3} {\mathbf k} \cdot 
\nabla \right ] \phi (t, \mathbf {x}) = 0\ ,
\label{eq3}
\end{equation}  
where $C_{2} = \omega / k^{2}$ and $C_{3} = \omega^2 / k^{2}
= C_1$ are arbitrary constants.  This means that there are two infinite 
sets of the second-order asymmetric equations.

The form of Eq. (\ref{eq2}) is the same as that of the Schr\"odinger 
equation [5], except for the presence of the coefficient $C_2$.  
Therefore, all equations of the same form as Eq. (\ref{eq2}) are 
called {\it Schr\"odinger-like equations.}  However, the equations 
given by Eq. (\ref{eq3}) with different coefficients $C_3$ are 
called {\it new asymmetric equations}, as originally named 
when the equation was first introduced [24].  It must also be 
noted that the constants $C_1$, $C_2$ and $C_3$ are expressed 
in terms of the eigenvalues, which label the irreps of $\mathcal {G}_e$.

In the previous work [24], it was shown that by using the de Broglie 
relationship [5], the coefficient $C_2$ expressed in terms of the labels 
of the irreps $\omega$ and $k$ can be evaluated, and it becomes the 
same as the coefficient in the Schr\"odinger equation of QM [5].  The 
obtained Schr\"odinger equation does not include any potentials, 
which means that it describes free quantum particles of ordinary 
matter.  Moreover, the coefficient $C_3$ of the new asymmetric 
equation was also evaluated, and the resulting equation was used 
to describe a quantum structure of dark matter particles [25].  In 
the following section, the coefficients $C_1$, $C_2$ and $C_3$ 
are evaluated in such a way that the resulting equations describe 
classical waves.    

\section{Wave equations for classical waves}

There are infinite sets of the symmetric (Eq. \ref{eq1}) and asymmetric 
(Eqs \ref{eq2} and \ref{eq3}) equations.  To select equations that describe
classical waves, the constants $C_1$, $C_2$ and $C_3$ must be expressed 
in terms of the wave frequency and wave vector as well as the wave speed.
This can be achieved by identifying the labels of the irreps $\omega$ and 
$k$ as the wave frequency and wave number, respectively, and introducing
the characteristic wave speed, $c_w = \omega / k$.  Then, Eq. (\ref{eq1}) 
becomes 
\begin{equation}
\left [ {{\partial^2} \over {\partial t^2}} - c^2_w \nabla^{2} \right ] 
\phi (t, \mathbf {x}) = 0\ , 
\label{eq4}
\end{equation}  
which is the well-known standard wave equation SWE [7-9].  Moreover, 
the Schr\"odinger-like and new asymmetric wave equations for classical 
waves can be written as 
\begin{equation}
\left [ i {{\partial} \over {\partial t}} + {{c_w^2} \over {\omega}} 
\nabla^{2} \right ] \phi (t, \mathbf {x}) = 0\ ,
\label{eq5}
\end{equation}  
and
\begin{equation}
\left [ {{\partial^2} \over {\partial t^2}} -  i  c_{w}^2 {\mathbf k} 
\cdot \nabla \right ] \phi (t, \mathbf {x}) = 0\ .
\label{eq6}
\end{equation}  
Observe that the obtained standard, Schr\"odinger-like, and new 
asymmetric wave equations are of different forms, and yet they 
can be used to describe free propagation of classical waves, as it 
is now demonstrated. 

If the wave speed $c_w$ is constant in the above wave equations,
then it is easy to verify that the solutions to the SWE given by Eq. 
(\ref{eq4}) are either  
\begin{equation}
\phi (t, \mathbf {x}) = A e^{-i (\omega t - {\mathbf k} \cdot 
{\mathbf r})} + B e^{-i (\omega t + {\mathbf k} \cdot 
{\mathbf r})}\ ,
\label{eq6a}
\end{equation}  
or
\begin{equation}
\phi (t, \mathbf {x}) = C e^{i (\omega t - {\mathbf k} \cdot 
{\mathbf r})} + D e^{i (\omega t + {\mathbf k} \cdot 
{\mathbf r})}\ , 
\label{eq6b}
\end{equation}  
where $A$, $B$, $C$ and $D$ are constants to be determined 
by specifying boundary conditions.  The solutions given by Eqs
(\ref{eq6a}) and (\ref{eq6b}) are equivalent and they reflect the 
fact that $\sqrt{-1} = \pm i$, which means that the choice of 
solutions is a matter of convention and it has no physical effect 
[7-9].  Moreover, the first and second solutions in Eq. (\ref{eq6a}) 
describe the forward and backward waves, respectively, and the 
same is true for Eq. (\ref{eq6b}). Substitution of any solution 
presented above into the SWE results in the dispersion relation 
$\omega^2 = k^2 c_w^2$, which verifies the choice of $C_1 
= c_w^2$ selected for Eq. (\ref{eq1}).  

For the Schr\"odinger-like wave equation given by Eq. (\ref{eq5}), 
the only solutions that describe classical waves are given by Eq. 
(\ref{eq6a}) as after substituting any of these two solutions into 
the equation, the dispersion relation $\omega^2 = k^2 c_w^2$
is obtained; this relation justifies the choice of $C_2 = c_w^2 / 
\omega$ in Eq. (\ref{eq2}).  However, the solutions given by 
Eq. (\ref{eq6b}) lead to the dispersion relation $\omega^2 = 
- k^2 c_w^2$, which does not represent waves, instead 
$\omega = - i k c_w$ describes exponentially decaying 
oscillations in the background medium.

The new asymmetric wave equation given by Eq. (\ref{eq5}) 
allows only for the solutions that can be written in the following 
form
\begin{equation}
\phi (t, \mathbf {x}) = A e^{i ({\mathbf k} \cdot {\mathbf r} -
\omega t)} + D e^{i ({\mathbf k} \cdot {\mathbf r} + \omega t)}\ , 
\label{eq6c}
\end{equation}  
where the first and second solutions represent the forward and 
backward propagating waves, respectively.  Substituting any of 
these two solutions into the new asymmetric wave equation
gives the dispersion relation $\omega^2 = k^2 c_w^2$, which
justifies the choice of $C_3 = c_w^2$ in Eq. (\ref{eq3}). The 
two other remaining solutions of Eqs (\ref{eq6a}) and (\ref{eq6b}) 
give the dispersion relation $\omega^2 = - k^2 c_w^2$, which 
does not describe waves but instead exponentially decaying 
oscillations with $\omega = + i k c_w$ in a medium where 
the waves propagate.

The presented results demonstrate that all three considered 
wave equations account for both the forward and backward 
waves, whose dispersion relations are the same, namely,  
$\omega^2 = k^2 c_w^2$, which allows expressing the 
coefficients $C_1$, $C_2$ and $C_3$ in terms of the wave 
speed $c_w$.  It is also shown that the standard wave
equation allows for two solutions identifed with either 
$+ i$ or $-i$.  However, the Schr\"odinger-like wave 
equation is limited to the solutions with $-i$, while the 
new asymmetric wave equation allows only for the 
solutions with $+i$, which means that these two wave 
equations are complementary.

The derived three wave equations are second-order, thus, 
they are local, which is one of the requirements for them 
to be fundamental.  Since the wave equations describe 
freely propagating waves, the requirement of gauge 
invariance [1] does not have to be considered.  However, 
it remains to be determined whether these equations have 
Lagrangians, and whether they are Galilean invariant or not. 

\section{Lagrangians for wave equations}

The Lagrangian formalism requires prior knowledge of a 
Lagrangian from which a dynamical equation is derived.  
Typically, the Lagrangians are presented without explaining 
their origin because there are no methods to derive them 
from first principles; however, for some systems, 
a Lagrangian can be constructed by accounting for the 
invariance of physical laws, the invariance of a physical 
system under consideration, and the structure of its 
equations (linear or nonlinear, driven or undriven, 
damped or undamped, etc.). Historically, most 
equations of physics were established first and only 
then their Lagrangians were found, often by guessing. 
 Once the Lagrangians are known, the process of finding 
the resulting equations is straightforward requiring 
substitution of these Lagrangians into the Euler-Lagrange 
(E-L) equation.   Despite some progress in deriving 
Lagrangians for physical systems described by ordinary 
differential equations (ODEs) (e.g., [26-31]), similar work 
for partial differential equations (PDEs) has only limited 
applications (e.g., [9,32,33]).  

Let $L (\phi, \partial_t \phi, \nabla \phi)$, where $\partial_t = \partial / 
\partial t$, be a Lagrangian that satisfies the E-L equation
\begin{equation}
{{\partial L} \over {\partial \phi}} - \partial_t \left ( {{\partial L} \over 
{\partial (\partial_t \phi)}} \right ) - \nabla \cdot \left ( {{\partial L} 
\over {\partial (\nabla \phi) }} \right ) = 0\ .
\label{eq7}
\end{equation}
Substituting $L (\phi, \partial_t \phi, \nabla \phi)$ into Eq. (\ref{eq7}) 
gives the required dynamical equation if, and only if, the Lagrangian 
is a priori known.  In case the equation is given first, its Lagrangian 
must be constructed in such a way that when substitued into Eq. 
(\ref{eq7}) the desired dynamical equation is obtained; this is the 
Lagrangian formalism.

Since the SWE given by Eq. (\ref{eq4}) is hyperbolic, its Lagrangian 
can be constructed [9,32] and the result is  
\begin{equation}
L_{swe} (\partial_t \phi, \nabla \phi) = {1 \over 2} \left [ c_w^{-2} 
(\partial_t \phi )^2 - ( \nabla \phi )^2 \right ]\ .
\label{eq8}
\end{equation}
It is easy to verify that substitution of the Lagrangian $L = L_{swe} 
(\partial_t \phi, \nabla \phi)$ into Eq. (\ref{eq7}) gives the required 
symmetric wave-like equation.

The Schr\"odinger-like wave equation given by Eq. (\ref{eq5}) is parabolic,
thus, its Lagrangian must be of a special form involving both $\phi$ and its 
complex conjugate $\phi^*$ [32].  The form of this Lagrangian is  
\begin{equation}
L_{Sch} (\phi, \phi^*, \partial_t \phi, \partial_t \phi^*, \nabla \phi,
\nabla \phi^*) = {i \over 2} \left ( \phi^* \partial_t \phi - \phi
\partial_t \phi^* \right ) - \frac{c_w^2}{\omega} ( \nabla \phi^* ) 
\cdot ( \nabla \phi )\ . 
\label{eq9}
\end{equation}
This Lagrangian gives the Schr\"odinger-like wave equation when 
substituted into the E-L equation for the variations in $\phi^*$.  On 
the other hand, the variations in $\phi$ lead to the complex conjugate 
Schr\"odinger-like wave equation, which becomes important when 
the probablity density $\vert \phi \vert^2$ is required.  However, 
in theories of classical waves, $\vert \phi \vert^2$ does not play 
any significant role as it does in QM [5]. 

To find the Lagrangian for the new asymmetric wave equation given by 
Eq. (\ref{eq6}), the Lagrangian for the Schr\"odinger-like wave equation 
must be modified as
\[
L_{asy} (\phi, \phi^*, \partial_t \phi, \partial_t \phi^*, \nabla \phi,
\nabla \phi^*) = (\partial_t \phi^*) (\partial_t \phi) +
\]
\begin{equation}
\hskip2.25in + {i \over 2} c_w^2 \left [ \phi^* ( \mathbf {k} \cdot 
\nabla \phi ) - \phi ( \mathbf {k} \cdot \nabla \phi^* ) \right ]\ .
\label{eq10}
\end{equation}
Then, this Lagrangian is substituted into the E-L equation
\begin{equation}
{{\partial L_{asy}} \over {\partial \phi^*}} - \partial_t \left ( {{\partial L_{asy}} 
\over {\partial (\partial_t \phi^*)}} \right ) - ( \mathbf {k} \cdot \nabla ) 
\cdot  \left ( {{\partial L_{asy}} \over {\partial (\mathbf {k} \cdot \nabla \phi^*)}} 
\right ) = 0\ ,
\label{eq11}
\end{equation}
and the new asymmetric wave equation (see Eq. \ref{eq6}) is obtained.

The presented results demonstrate that Lagrangians exist for the 
symmetric and asymmetric wave equations, and that these equations 
are local.  Therefore, the last requirement for a wave equation to 
be called fundamental is its Galilean invariance, which is now 
investigated.

\section{Galilean invariance of wave equations}

\subsection{Known fundamental equations of nonrelativistic physics}

Let $S$ and $S^{\prime}$ be two inertial frames moving with respect 
to each other with the velocity $\mathbf {v}$  = const, which allows 
writing a boost as $\mathbf {x} = \mathbf {x}^{\prime} + \mathbf {v} 
t^{\ \prime}$ with $t^{\ \prime}\ =\ t$.  Then, the Galilean metric in 
space is $ds^2 = d \mathbf {x} \cdot d \mathbf {x} = dx^2 + dy^2 + 
dz^2$ with $ds^2 = ds^{\prime 2}$, and in time $dt^2 = dt^{\prime 2}$.  
By performing the Galilean transformations (translations in space and time, 
rotations and boosts) that form the Galilean group of the metric [11], or 
the extended Galilean group $\mathcal {G}_e$ [10], Galilean invariance 
of the metrics can be verified.  The invariance means that the forms of 
the metrics remain the same for all Galilean observers. 

Similarly, for a dynamical equation to be Galilean invariant, it is required 
that the form of the equation remains the same in all inertial frames; 
this means that the coefficients of this equation must also be the same
in all inertial frames.  With the equation retaining its form, solutions of 
this equation are also the same for all Galilean observers.  The simplest 
example is the second-order ODE describing the law of inertia, whose 
invariance with respect to all transformations that form the Galilean 
group of the metrics is well-known [3,11]. It is also known that the 
Lagrangian of the law of inertia is not Galilean invariant [2,3,11]; 
however, it was recently shown that Galilean invariance of the 
Lagrangian can be restored by using the so-called null Lagrangians 
[34].  Thus, the law of inertia is a fundamental equation of 
classical mechanics. 

As shown above, space and time in Galilean relativity are separated and 
obey different metrics.  Therefore, for dynamical equations to be
Galilean invariant they must be asymmetric in time and space derivatives.
The ODE describing the law of inertia is asymmetric as it does not have 
any space derivative.  However, among the wave equations obtained in 
this paper and given by Eqs. (\ref{eq4}), (\ref{eq5}) and (\ref{eq6}),
the SWE is symmetric and the two other wave equations are asymmetric.
As a result, {\it the SWE is not Galilean invariant} and, thus, {\it it is not 
fundamental.} In other words, for classical particles, the law of inertia is
the fundamental equation, but there is no corresponding fundamental 
equation for classical waves; the main objective of this paper is to find 
such an equation and apply it to wave theories. 

The Sch\"odinger equation of QM is asymmetric and its Galilean 
invariance is well-known, requiring a phase factor, whose form is 
frame-dependent [5,10,16-18].  The existence of this phase factor 
makes the wavefunction to be different for each Galilean observer, 
which may imply that the equation is not Galilean invariant. However, 
the presence of the phase factor in the solutions does not violate 
Galilean invariance because in QM only the square of the absolute 
value of the wavefunction is the measureable quantity, and this 
quantity remains the same for all Galilean observers.  Thus, the 
Schr\"odinger equation for free quantum particles is a fundamental 
equation of QM.  Similarly, the  L\'evy-Leblond equation for its spinor 
wavefunction is Galilean invariant and a fundamental equation of 
QM [6,10].  Moreover, the Lagrangians for the Schr\"odinger and 
 L\'evy-Leblond equations are Galilean invariant.

Having demonstrated that the law of inertia and the Schr\"odinger 
and  L\'evy-Leblond equations are the fundamental equations of 
nonrelativistic physics, and that the SWE cannot be a fundamental 
equation for classical waves, it remains now to determine whether 
the Schr\"odinger-like and new asymmetric equations are fundamental. 

\subsection{Schr\"odinger-like wave equation}

Applying the Galilean transformations to Eq. (\ref{eq5}), the transformed 
Schr\"odinger-like wave equation can be written as 
\begin{equation}
\left [ i {{\partial} \over {\partial t^{\prime}}} + {{c_w^{\prime 2}} 
\over {\omega^{\prime}}} \nabla^{\prime 2} \right ] \phi^{\prime} 
(t^{\prime}, \mathbf {x^{\prime}}) = 0\ ,
\label{eq12}
\end{equation}  
where the original and transformed wavefunctions are related by
\begin{equation} 
\phi (t, \mathbf {x}) = \phi (t^{\prime}, \mathbf{x^{\prime}} + 
\mathbf {v} t^{\prime}) = \phi^{\prime} (t^{\prime}, \mathbf{x^{\prime}})\ 
e^{i \eta (t^{\prime}, \mathbf{x^{\prime}})}\ ,
\label{eq13}
\end{equation}
with the phase factor given by
\begin{equation}
\eta (t^{\prime}, {\mathbf {x}^{\prime}}) = \frac{\omega^{\prime}}
{2 c_w^{\prime\ 2}} \left ( \mathbf {v} \cdot \mathbf {x}^{\prime} + 
v^2 t^{\prime} / 2 \right )\ .
\label{eq14}
\end{equation}

For the obtained transformed Schr\"odinger-like wave equation to 
be Galilean invariant, it is also required that $c_w^2 / \omega = 
c_w^{\prime 2} / \omega^{\prime}$.   This condition is satisfied 
when
\begin{equation}
\mathbf {k}^{\prime} = \mathbf {k} - {{\omega} \over {2 c_w^2}} 
\mathbf {v}\ , 
\label{eq15}
\end{equation}
and
\begin{equation}
\omega^{\prime} = \omega  \left ( 1 + \frac{v^2}{4 c_w^2} \right) 
- \mathbf {k} \cdot \mathbf {v}\ . 
\label{eq16}
\end{equation}

The above results demonstrate that the Schr\"odinger-like wave equation 
preserves its form in all inertial frames if, and only if, the wavefunction
transforms according to Eq. (\ref{eq13}), and Eqs. (\ref{eq15}) and 
(\ref{eq16}) are satisfied.  The existence of the phase factor given by 
Eq. (\ref{eq14}), which is a frame-dependent quantity, is well-known 
and its presence does not violate Galilean invariance of the Sch\"odinger 
equation in QM because of its requirement that only $\vert \phi (t, \mathbf 
{x}) \vert^2 = \vert \phi^{\prime} (t^{\prime}, \mathbf{x^{\prime}}) 
\vert^2$ must be valid for all Galilean observers [5,10,16-18]. 

For classical waves, the wavefunction $\phi (t, \mathbf {x})$ represents
one of the physical variables describing a wave; thus, to get the same 
wave description by all Galilean observers, the solutions for $\phi (t, 
\mathbf {x})$ and $\phi^{\prime} (t^{\prime}, \mathbf{x^{\prime}})$
must be the same in all inertial frames.  However, they are not because 
of the presence of the phase factor (see Eq. \ref{eq13}) that is different 
in different inertial frames.  As a result, Galilean observers describe waves 
differently in their inertial frames, which means that the Schr\"odinger-like 
equation for classical waves is not Galilean invariant and therefore it is not 
fundamental.  

\subsection{New asymmetric wave equation} 
   
Let $\phi (t, \mathbf {x})$ be the wavefunction of Eq. (\ref{eq6}) and 
$\phi ({\mathbf {x}^{\prime}}, t^{\prime})$ be the transformed 
wavefunction.  After performing the Galilean transformations, Eq. 
(\ref{eq6}) becomes
\begin{equation}
\left [ {{\partial^2} \over {\partial t^{\prime 2}}} - i c^{\prime 2}_{w} 
\mathbf {k^{\prime}} \cdot \nabla^{\prime} \right ] \phi (t^{\prime}, 
\mathbf {x^{\prime}}) = \left [ 2 ( {\mathbf v} \cdot \nabla^{\prime} ) 
{{\partial} \over {\partial t^{\prime}}} - ( {\mathbf v} \cdot 
\nabla^{\prime} )^2 \right ] \phi (t^{\prime}, \mathbf {x^{\prime}})\ .
\label{eq17}
\end{equation}  
Comparison of this equation to Eq. (\ref{eq6}) shows that its 
LHS is of the same form as the new asymmetric wave equation 
if, and only if, the RHS is zero.  Let $\phi^{\prime} (t^{\prime}, 
{\mathbf {x}^{\prime}})$ be the wavefunction that satisfies 
the RHS of Eq. (\ref{eq17}).  As already demonstrated [24], 
the solution to the RHS of Eq. (\ref{eq17}) is any function 
$\phi^{\prime} (t^{\prime}, {\mathbf {x}^{\prime}}) = 
\phi^{\prime} ({\mathbf {r}^{\prime}})$, where 
${\mathbf {r}^{\prime}} = {\mathbf {x}^{\prime}} + 
{\mathbf {v}} t^{\prime} / 2$.  

Then, with $\phi (t^{\prime}, {\mathbf {x}^{\prime}}) 
= \phi^{\prime} (t^{\prime}, {\mathbf {x}^{\prime}}) 
= \phi^{\prime} ({\mathbf {r}^{\prime}})$, the LHS of
 Eq. (\ref{eq17}) can be written as 
\begin{equation}
\left [ {{d^2} \over {d {\mathbf r^{\prime 2}}}} - i \left ( 
\frac{2 c_{w}^{\prime}} {v}\right)^2 {\mathbf {k^{\prime}}} 
\cdot {{d} \over {d {\mathbf r^{\prime}}}}\right ] \phi^{\prime} 
(\mathbf {r^{\prime}})= 0\ .
\label{eq18}
\end{equation}
Using the Galilean transformations, ${\mathbf r^{\prime}} = 
{\mathbf {x}} - {\mathbf {v}} t / 2 \equiv {\mathbf {r}}$,
and Eq. (\ref{eq6}) becomes  
\begin{equation}
\left [ {{d^2} \over {d {\mathbf r^2}}} - i \left ( \frac{2 c_{w}} 
{v}\right)^2 {\mathbf {k}} \cdot {{d} \over {d {\mathbf r}}}
\right ] \phi (\mathbf {r})= 0\ ,
\label{eq19}
\end{equation}
where $\phi (\mathbf {r}) = \phi (t, {\mathbf {x}}) = \phi
 (t^{\prime}, {\mathbf {x}^{\prime}}) = \phi^{\prime}
 (t^{\prime}, {\mathbf {x}^{\prime}}) = \phi^{\prime} 
(\mathbf {r^{\prime}})$.  Then, Eqs (\ref{eq18}) and 
(\ref{eq19}) are of the same form if, and only if, 
$c_w^{\prime} = c_w$ and $\mathbf {k^{\prime}} = 
\mathbf {k}$.  To show that these conditions are valid, 
let the phase of a wave in the inertial frame $S^{\prime}$ 
be given by 
\begin{equation}
\mathbf {k^{\prime}} \cdot \mathbf {r^{\prime}} = \mathbf {k^{\prime}}
\cdot \mathbf {x^{\prime}} + \frac{1}{2} \left ( \mathbf {k^{\prime}} 
\cdot \mathbf {v} \right ) t^{\prime} = {\mathbf {k^{\prime}}} \cdot 
{\mathbf {x^{\prime}}} + \frac{1}{2} \left ( \frac{\omega^{\prime}}
{c_w^{\prime}} \right ) ( {\hat {\mathbf k^{\prime}}} \cdot 
{\hat {\mathbf {v}}} ) v t^{\prime}\ ,
\label{eq20}
\end{equation}
where $\hat {\mathbf k}$ and $\hat {\mathbf v}$ are unit vectors corresponding 
to $\mathbf k$ and $\mathbf v$, respectively. It must also be noted that the 
dispersion relation $\omega^{\prime} = k^{\prime} c_w^{\prime}$ was used 
to obtain the wave phase.   

After the Galilean transformations, Eq. (\ref{eq20}) reduces to
\begin{equation}
\mathbf {k^{\prime}} \cdot \mathbf {r^{\prime}} = {\mathbf {k^{\prime}}} 
\cdot {\mathbf {x}} - \frac{1}{2} \left ( \frac{\omega^{\prime}} {c_w^{\prime}} 
\right ) ( {\hat {\mathbf k^{\prime}}} \cdot {\hat {\mathbf {v}}} )\ v t\ ,
\label{eq21a}
\end{equation}
which represents the forward waves in an inertial frame $S^{\prime}$ (see Eq. 
\ref{eq6c}).  However, since $( {\hat {\mathbf k^{\prime}}} \cdot {\hat 
{\mathbf {v}}} ) = \cos \theta^{\prime}$ can be either positive or negative, 
$\mathbf {k^{\prime}} \cdot \mathbf {r^{\prime}}$ may also describe the 
backward waves if $\cos \theta^{\prime} < 0$.  This can be fixed by writing 
\begin{equation}
\mathbf {k^{\prime}} \cdot \mathbf {r^{\prime}} = {\mathbf {k^{\prime}}} 
\cdot {\mathbf {x}} \pm \frac{1}{2} \left ( \frac{\omega^{\prime}} {c_w^{\prime}} 
\right ) \vert {\hat {\mathbf k^{\prime}}} \cdot {\hat {\mathbf {v}}} \vert\ v t\ ,
\label{eq21b}
\end{equation}
where the $+$ and $-$ signs correspond to the backward and forward waves, 
respectively.

On the other hand, the wave phase in an inertial frame $S$ is
\begin{equation}
\mathbf {k} \cdot \mathbf {r} = {\mathbf {k}} \cdot {\mathbf {x}} \pm  
\frac{1}{2} \left ( \frac{\omega} {c_w} \right ) \vert {\hat {\mathbf k}} 
\cdot {\hat {\mathbf {v}}} \vert \ v t\ ,
\label{eq22}
\end{equation}
with the $+$ and $-$ signs corresponding to the backward and forward 
waves, respectively.

The requirement of Galilean invariance is that the wave phases are the 
same in all inertial frames, which means that ${\mathbf k^{\prime}} 
\cdot {\mathbf r^{\prime}} = {\mathbf k} \cdot {\mathbf r}$. Hence, 
\begin{equation}
( {\mathbf {k^{\prime}}} - {\mathbf {k}} ) \cdot {\mathbf {x}} \pm
\frac{1}{2} \left [ \frac{\omega^{\prime}} {c_w^{\prime}} \vert 
{\hat {\mathbf k^{\prime}}} \cdot {\hat {\mathbf {v}}} \vert -
\frac{\omega}{c_w} \vert {\hat {\mathbf {k}}} \cdot {\hat {\mathbf 
{v}}} \vert \right ] v t = 0\ ,
\label{eq23}
\end{equation}
which is only satisfied when $\mathbf {k^{\prime}} = \mathbf {k}$ and
$\omega^{\prime} / c_w^{\prime} = \omega / c_w$.  

With $\mathbf {k^{\prime}} = \mathbf {k}$ = const (see Eq. \ref{eq1b}), 
Eqs (\ref{eq17}) and (\ref{eq18}) can be written in the following form
\begin{equation}
{{d} \over {d ({\mathbf k^{\prime}} \cdot {\mathbf r^{\prime}})}} 
\left [ {{d} \over {d ({\mathbf k^{\prime}} \cdot {\mathbf 
r^{\prime}})}} - i \left ( \frac{2 c_{w}^{\prime}}{v}\right)^2 
\right ] \phi^{\prime} ({\mathbf k^{\prime}} \cdot 
{\mathbf r^{\prime}})= 0\ ,
\label{eq24}
\end{equation}
and 
\begin{equation}
{{d} \over {d ({\mathbf k} \cdot {\mathbf r})}} \left [ {{d} \over 
{d ({\mathbf k} \cdot {\mathbf r})}} - i \left ( \frac{2 c_{w}}{v}\right)^2 
\right ] \phi ({\mathbf k} \cdot {\mathbf r})= 0\ .
\label{eq25}
\end{equation}
Since ${\mathbf k^{\prime}} \cdot {\mathbf r^{\prime}} = {\mathbf k} 
\cdot {\mathbf r}$, $\phi^{\prime} ({\mathbf k^{\prime}} \cdot 
{\mathbf r^{\prime}}) = \phi ({\mathbf k} \cdot {\mathbf r})$ and 
$c_w^{\prime} = c_w$, the above equations are of the same form, 
they are Galilean invariant, and this invariance does not require 
any phase factor.  However, the form of Eq. (\ref{eq25}) is very 
different from that of the original new asymmetric equation given 
by Eq. (\ref{eq6}), which means that in order for this equation to 
be fundamental, the existence of its Lagrangian must be established.

\subsection{Fundamental wave equation for classical waves}

The Galilean invariant equation (Eq. \ref{eq25}) is an ordinary
differential equation, whose Lagrangian can be found by one of the 
methods previously developed for ODEs (e.g., [26-31]).  The 
Lagrangian for Eq. (\ref{eq25}) can be written as 
\begin{equation}
L_{as} (d_{kr} \phi , {\mathbf k} \cdot {\mathbf r}) = {1 \over 2} 
\left [ d_{kr} \phi ({\mathbf k} \cdot {\mathbf r}) \right ]^2\ 
e^{- 4 i ({\mathbf k} \cdot {\mathbf r}) c^2_w / v^2}\ ,
\label{eq26}
\end{equation}
where $d_{kr} = d/d({\mathbf k} \cdot {\mathbf r})$. The derived 
Lagrangian depends on the wave phase ${\mathbf k} \cdot {\mathbf r}$ 
that involves both $\mathbf {x}$ and $t$.  

In CM, the dependence of Lagrangians on $t$ implies that the total 
energy of a dynamical system is not conserved and, as a result, the 
energy function must be calculated [3,4].  For physical systems with 
their Lagrangians explicitly time-dependent, the exponentially decaying 
or increasing terms are present, like in the well-known Caldirola-Kanai 
Lagrangian [35,36], originally written for the Bateman oscillator [37,38]. 
However, the Lagrangian given by Eq. (\ref{eq26}) is of a different form 
as its exponential term is periodic in ${\mathbf k} \cdot {\mathbf r}$ 
instead.  Since the first term on the RHS in Eq. (\ref{eq26}) represents 
the wave kinetic energy, the exponential term shows that this energy is 
required to oscillate in time and space in the Lagrangian, so that the 
correct wave equation is obtained.  This is a new phenomenon in 
classical waves and, thus, $L_{new} (d_{kr} \phi , {\mathbf k} 
\cdot {\mathbf r}) $ forms a separate class among all Lagrangians 
known in physics.

To demonstrate that the Lagrangian $L_{as} (d_z \phi , z)$ is 
Galilean invariant, the Galilean transformations are applied and 
the following transformed Lagrangian is found
\begin{equation}
L_{as}^{\prime} (d_{kr}^{\prime} \phi^{\prime} , {\mathbf 
k^{\prime}} \cdot {\mathbf r^{\prime}}) = {1 \over 2} \left [ 
d_{kr}^{\prime} \phi^{\prime} ({\mathbf k^{\prime}} \cdot 
{\mathbf r^{\prime}}) \right ]^2\ e^{- 4 i ({\mathbf k^{\prime}} 
\cdot {\mathbf r^{\prime}}) c_w^{\prime 2} / v^2}\ ,
\label{eq27}
\end{equation}
where $d_{kr}^{\prime} = d/d({\mathbf k^{\prime}} \cdot 
{\mathbf r^{\prime}})$. The transformed Lagrangian is of the
same form as the original one given by Eq. (\ref{eq26}) because
${\mathbf k} \cdot {\mathbf r} = {\mathbf k^{\prime}} \cdot 
{\mathbf r^{\prime}}$, $\phi ({\mathbf k} \cdot {\mathbf r}) 
= \phi^{\prime} ({\mathbf {r}^{\prime}} \cdot {\mathbf 
k^{\prime}})$ and $c^2_w / v^2 = c^{\prime\ 2}_w / v^2$.  
Therefore, the Lagrangian $L_{as} (d_{kr} \phi , {\mathbf k} 
\cdot {\mathbf r})$ is Galilean invariant.

After substituting the Lagrangian given by Eq. (\ref{eq26}) into 
the E-L equation 
\begin{equation}
 d_{kr} \left ( \frac{d L_{as}} {d (d_{kr} \phi)} \right ) - 
\frac{d L_{as}}{d \phi} = 0\ ,
\label{eq27}
\end{equation}
the following equation is obtained
\begin{equation}
\left [ {{d^2 \phi} \over {d ({\mathbf k} 
\cdot {\mathbf r})^2}} - i \left ( \frac{2 c_{w}}{v}\right)^2 {{d \phi} 
\over {d ({\mathbf k} \cdot {\mathbf r})}} \right ] e^{- 4 i ({\mathbf k} 
\cdot {\mathbf r}) c^2_w / v^2}= 0\ .
\label{eq28}
\end{equation}
Since $e^{- 4 i ({\mathbf k} \cdot {\mathbf r}) c^2_w / v^2} \neq 0$,
the terms in the square brackets must be zero, which gives Eq. (\ref{eq25}).
This shows that in addition to be local and Galilean invariant, Eq. (\ref{eq25}) 
can also be derived from the Lagrangian given by Eq. (\ref{eq26}).  With 
its Lagrangian known and Galilean invariance of the Lagrangian verified, 
Eq. (\ref{eq25}) is the new fundamental asymmetric wave equation or 
simply the {\it fundamental wave equation} (FWE).  By being fundamental, 
the FWE gives the most comprehensive description of free classical waves, 
as it accounts for the Doppler effect, the forward and backward waves, 
and makes the wave speed to be the same in all inertial frames.  

The wave speed $c_w$ is constant for all Galilean observers, and 
since $v$ = const, the coefficient $4 c_w^2 / v^2$ = const.  This 
is an interesting result. It shows that this coefficient plays similar 
role for classical waves in Galilean relativity as the speed of light 
$c$ plays in Special Theory of Relativity (STR) for electromagnetic 
(EM) waves.  However, while $c = {\rm const}$ is the basic principle 
of Nature and the foundation of STR, the coefficient $4 c_w^2 / 
v^2$ = const is the necessary condition for Galilean invariance, 
and its validity is guaranteed by the existence of the FWE, 
and by selecting the wave phase as the variable representing the 
waves.

Thus, the main result of this paper is that classical waves 
propagating with speeds $c_w << c$ may 'mimic' the behavior 
of EM waves in STR when they are described by the FWE.  For 
this reason, it is suggested that these waves be called the {\it 
basic classical waves} in Galilean Relativity.

\section{Applications to acoustic wave propagation}

\subsection{Freely propagating acoustic waves}

Acoustic waves propagate freely in uniform media and the solutions 
of the SWE that describe such propagation are given by Eqs. (\ref{eq6a}) 
and (\ref{eq6b}), with the wave frequency $\omega$ and the wave 
vector $\mathbf {k}$ being frame-dependent (the Doppler effect);
this means that Galilean observers see plane waves with their 
frequencies and wave vectors being different in their respective 
inertial frames moving with constant velocity $\mathbf {v}$.  
Therefore, the SWE is not Galilean invariant, and thus it is not 
fundamental.

Finding the solutions to the FWE given by Eq. (\ref{eq25}) is 
straightforward.  After two integrations, it yields
\begin{equation}
\phi ({\mathbf k} \cdot {\mathbf r}) = c_1 e^{i \theta_s} + c_2\ ,
\label{eq29}
\end{equation}
where $c_1$ and $c_2$ are integration constants, and the phase of 
the acoustic wave is   
\begin{equation}
\theta_s \equiv \left ( \frac{2 c_{s}}{v} \right )^2 ({\mathbf k} \cdot 
{\mathbf r}) = \left ( \frac{2 c_{s}}{v} \right )^2 \left [ {\mathbf k} 
\cdot {\mathbf x} \pm \frac{1}{2} \left ( \frac{v} {2 c_{s}} \right ) 
\vert {\mathbf {\hat k}} \cdot {\mathbf {\hat v}}\vert\ \omega t 
\right ]\ ,
\label{eq30}
\end{equation}
where $c_w \equiv c_s$ is the speed of sound.  The solution for 
$\phi ({\mathbf k} \cdot {\mathbf r})$ describes both the forward 
and backward propagating acoustic waves (see Eq. \ref{eq6c}). The 
conditions ${\mathbf k} \cdot {\mathbf r} = {\mathbf k}^{\prime} 
\cdot {\mathbf r}^{\prime}$ and $(2 c_s / v)^2 = (2 c_s^{\prime} 
/ v)^2$ guarantee that the solution is the same in all inertial frames 
and that it accounts for the Doppler effect.  Thus, the above solution 
shows that its description of acoustic waves freely propagating in an 
uniform medium is much more comprehensive than that given by 
the SWE.

In the next section, the assumption of uniform media is removed 
and a gradient of density is included, making the background 
medium stratified.

\subsection{Lamb's cutoff frequency}

In his original work, Lamb [39-41] considered acoustic waves 
propagating in the $z$-direction in the background medium with 
gravity $\vec g = - g \hat z$ and density gradient $\rho_0 (z) 
= \rho_{00} \exp (- z / H)$, where $\rho_{00}$ is the gas density 
at the height $z=0$, and $H = c_s^2 / \gamma g$ is the density 
scale height, with $\gamma$ denoting the ratio of specific heats. 
In his model, the background gas pressure $p_0$ and gas density 
$\rho_0$ vary with height $z$; however, the temperature $T_0$ 
remains constant. As a result, $H$ = const and $c_s$ = const. 

This stratified but otherwise isothermal medium is often referred to 
as an {\it isothermal atmosphere}, and acoustic waves in this atmosphere 
are described by the following variables: velocity $u (t,z)$, pressure 
$p (t,z)$ and density $\rho (t,z)$ perturbations.  The resulting acoustic
wave equation (AWE) is derived for the transformed wave variables 
$u_1 (t,z) = u (t,z) \rho_{0}^{1/2}$, $p_1 (t,z) = p (t,z) 
\rho_{0}^{-1/2}$ and $\rho_1 (t,z) = \rho (t,z) \rho_{0}^{-1/2}$ 
using the hydrodynamic equations [40-42,46].  The resulting wave 
equation can be written as 
\begin{equation}
\left [ {\partial^2 \over \partial t^2} - c_s^2\ {\partial^2 \over 
\partial z^2} + \Omega_{ac}^2 \right ] [ u_1 (t,z), p_1 (t,z), 
\rho_1 (t,z)]\ =\ 0\ .
\label{eq30}
\end{equation}
where the speed of sound is $c_s = [\gamma p_0 (z) / \rho_0 (z)]^{1/2} 
= [\gamma R T_0 / \mu]^{1/2}$, while the acoustic cutoff frequency 
$\Omega_{ac} = c_s / 2 H = \gamma g / 2c_s$ remains constant in 
the entire isothermal atmosphere [39-41,42,46]. The Lamb cutoff 
frequency describes the effects of the atmospheric density gradient 
on the acoustic wave propagation, and it is used to determine the 
wave propagation conditions (see Sec. 6.4).  Note also that the 
form of the wave equation is the same for each wave variable 
in an isothermal atmosphere.

The fact that the form of the derived AWE remains the same at every 
height in an isothermal atmosphere is well-known and it was first 
shown by Lamb [39-41].  However, different inertial observers see 
the waves differently, namely, with their different characteristic 
speeds, frequencies, and wave vectors.  Different waves seen in 
different inertial frames means that the theory of waves based 
on the AWE is not fundamental because it is not the same for all 
Galilean observers.

In numerous studies of propagation of acoustic waves that followed Lamb's 
work, different aspects of the wave propagation were investigated by using 
methods based on either global and local dispersion relations, or the WKB 
approximation, or finding analytical solutions to acoustic wave equations 
for special cases [43-45].  A method to determine the cutoff frequency 
for linear and adiabatic acoustic waves propagating in non-isothermal 
media without gravity was also developed [46] based on transformations 
of wave variables that lead to standard wave equations, and using the 
oscillation theorem to determine the turning point frequencies. Physical 
arguments are used to select the largest of these frequencies as the 
Lamb cutoff frequency.  In this paper, the Lamb cutoff frequency is 
obtained for the new fundamental wave equation.

\subsection{Fundamental wave equation and Lamb's cutoff frequency}

The acoustic wave equation given by Eq. (\ref{eq30}) is obtained 
from the hydrodynamic equations.  It is easy to show that neither 
the Schr\"odinger-like wave equation nor the new asymmetric wave 
equation can be derived using only the hydrodynamic equations.  
However, both wave equations can be derived from the hydrodynamic 
equations if, and only if, these equations are supplemented by the 
eigenvalue equations. Specifically, the Schr\"odinger-like wave 
equation is obtained when the eigenvalue equation given by 
Eq. (\ref{eq1a}) is applied to Eq. (\ref{eq30}).  However, the 
Schr\"odinger-like wave equation is not fundamental (see Section
5.2), therefore, the equation will not be further considered here.

Instead, the new asymmetric equation given by Eq. (\ref{eq5}) 
is considered.  By applying the eigenvalue equation given by Eq. 
(\ref{eq1b}) to Eq. (\ref{eq30}), the following equation is 
obtained 
\begin{equation}
\left [ {\partial^2 \over \partial t^2} - i k {c_s^2} {\partial \over 
\partial z} + \Omega_{ac}^2 \right ] \phi (t,z)\ =\ 0\ .
\label{eq31}
\end{equation}
For the considered acoustic wave propagation along the $z$-axis,
the label $\mathbf{k}$ of the irreps of $\mathcal {G}_e$ is 
identified with the wave vector and $k = {\mathbf k} \cdot \hat z$.
In addition, the wavefunction $\phi (t,z)$ represents one of the 
acoustic wave variables in Eq. (\ref{eq30}).  

The results presented in Section 5.3 demonstrate that the 
new asymmetric wave equation can be converted into a form 
that is Galilean invariant (see Eq. \ref{eq25}).  Applying the 
results to Eq. (\ref{eq31}), the resulting wave equation is    
\begin{equation}
\left [ \frac{d^2 }{d{\chi}^2} - i \left ( \frac{2 c_{s}}{v}
\right )^2 \frac{d}{d \chi} + \left ( \frac{2 \Omega_{ac}}
{k v} \right )^2 \right ] \phi (\chi) = 0\ ,
\label{eq32}
\end{equation}
where $\chi = {\mathbf k} \cdot {\mathbf r} = k (z \pm 
\vert \hat{z} \cdot \hat {v} \vert v t / 2)$.  Since 
$\Omega_{ac} = \gamma g / 2 c_s$, with $c_s = 
c_s^{\prime}$, $k = k^{\prime}$, and with $\gamma$ 
and  $g$ being the same in all inertial frames, Eq. 
(\ref{eq32}) and its solutions are the same for all 
Galilean observers; this means that the derived equation 
is the FWE for the considered acoustic waves.  The 
obtained FWE describes the effects of an isothermal 
atmosphere on the acoustic wave propagation.  Thus, 
Eq. (\ref{eq32}) generalizes Eq. (\ref{eq25}), which 
describes only freely propagating acoustic waves in a 
medium without any gradients.  

As a result of the Galilean transformations, the term that 
represents Lamb's cutoff frequency is now modified by the 
factor $2 / kv$, which describes the effects of moving inertial 
frames on the cutoff; these effects are more prominent for 
smaller velocities $v$ and wavevectors $k$.  All the presented 
results are valid for $v > 0$ (see Section 5.1), which means 
that if $S^{\prime}$ moves with respect to $S$ with velocity
$\mathbf {v}$, then $S$ moves with respect to $S^{\prime}$
with velocity $- \mathbf {v}$.  In case, there is only one 
stationary inertial frame with $v = 0$, this frame must be 
treated separately by using Eq. (\ref{eq6}) that is not Galilean 
transformed.  It is also important to point out that any Galilean 
observer may boost its inertial frame to the wave frame by 
setting $v = c_s$.

\subsection{Conditions for acoustic wave propagation}

As originally demonstrated by Lamb [39-41], the frequency 
$\Omega_{ac}$ uniquely determines whether acoustic waves
in an isothermal atmosphere are propagating or evanescent.
Since $\Omega_{ac}$ = const in the isothermal atmosphere,
after making the Fourier transforms in time and space, the 
AWE (Eq. \ref{eq30}) gives the global dispersion relation:
$(\omega^2 - \Omega_{ac}^2) = k^2 c_s^2$, where 
$\omega$ is the wave frequency and $k = \mathbf {k} 
\cdot \hat z$ is the wave vector. The obtained dispersion 
relation is valid in one selected inertial frame, which is 
called {\it stationary}.  In this frame, the waves are 
propagating when $\omega > \Omega_{ac}$ and $k$ 
is real, and they are non-propagating (evanescent) when 
either $\omega = \Omega_{ac}$ with $k = 0$ or $\omega 
< \Omega_{ac}$ with $k$ being imaginary.  

When a Galilean observer moves with velocity $v$ with respect to 
the stationary frame, then the wave frequency (the Doppler effect), 
wave vector and characteristic wave speed change, which means 
that $(\omega^{\prime 2} - \Omega_{ac}^{\prime 2}) = k^{\prime 2} 
c_s^{\prime 2}$; the dispersion relation preserves its form but 
the values of the wave parameters change from one inertial frame 
to another. With $\Omega_{ac} \neq \Omega_{ac}^{\prime}$, 
the acoustic cutoff frequency is different in different inertial frames.  
  
The same conditions for wave propagation are obtained when the 
Fourier transforms in time and space are performed in the new 
asymmetric wave equation  (Eq. \ref{eq31}), and the result 
is $(\omega^2 - \Omega_{ac}^2) = k^2 c_s^2$, which is the 
same as the dispersion relation obtained for the AWE.  Thus, the 
conditions for the wave propagation are also the same.  However,
neither the AWE given by Eq. (\ref{eq30}) nor the new asymmetric 
wave equation given by Eq. (\ref{eq31}) is fundamental.  The only 
FWE is given by Eq. (\ref{eq32}).  The conditions for wave propagation 
resulting from this equation are now determined and discussed. 

To find the conditions for the acoustic wave propagation in an 
isothermal atmosphere, the FWE given by Eq. (\ref{eq32}) 
must be solved.  The obtained solutions $\phi_1 (\chi)$ 
and $\phi_2 (\chi)$ are  
\begin{equation}
\phi_{1,2} (\chi) = \exp{ \left [\frac{i}{2} \left ( \frac{2 c_{s}}{v}
\right )^2 \left ( 1 \pm \sqrt{1 +  \left ( \frac{v}{c_{s}} \right )^2 
\frac{\Omega_{ac}^2}{\omega^2 - \Omega_{ac}^2}} \right ) 
\chi \right ]}\ ,
\label{eq33}
\end{equation}
and their superposition gives the general solution $\phi (\chi) = 
C_1 \phi_1 (\chi) + C_2 \phi_2 (\chi)$; note that the dispersion 
relation $k^2 c_s^2 = (\omega^2 - \Omega_{ac}^2)$ was used
to derive Eq. (\ref{eq33}).  Using the dispersion relation,  the wave 
phase $\chi = {\mathbf k} \cdot {\mathbf r} = k (z \pm \vert \hat{z} 
\cdot \hat {v} \vert v t / 2)$ can be written as 
\begin{equation}
\chi = \left ( \frac{z}{c_s} \pm \frac{1}{2} \frac{v}{c_s}
\vert \hat{z} \cdot \hat {v} \vert t \right ) 
\sqrt{\omega^2 - \Omega_{ac}^2}\ ,
\label{eq34}
\end{equation}
which allows writing the solutions given by Eq. (\ref{eq33}) in the 
following form
\begin{equation}
\theta_{1,2} (t, z) = \frac{c_{s}}{v} \left [ 
\sqrt{\omega^2 - \Omega_{ac}^2} \pm \sqrt{\omega^2 - \left ( 1 - 
\frac{v^2} {c_{s}^2} \right ) \Omega_{ac}^2} \right ] \left ( 
\frac{2z}{v} \pm \vert \hat{z} \cdot \hat {v} \vert  t \right )
\label{eq35}
\end{equation}

and the general solution becomes
\begin{equation}
\phi (t, z) = C_1 e^{i \theta_1(t, z)} + C_2 e^{i \theta_2 (t, z)}\ .
\label{eq36}
\end{equation}
This solution and its wave phases are now used to determine 
the conditions for the propagation of acoustic waves in an 
isothermal atmosphere.

The general solution given by Eq. (\ref{eq35}) shows that any 
real $\theta_1(t, z)$ and $\theta_2(t, z)$ describe propagating 
waves.  On the other hand, imaginary wave phases make the 
general solution exponentially decaying, which corresponds to 
non-propagating (or evanescent) waves.  There are several 
cases of interest that are now considered.  If $\omega >> 
\Omega_{ac}$, then the wave phases are $\theta_1 (t, z) 
= (2z / v \pm \vert \hat{z} \cdot \hat {v} \vert t) \omega$ 
and $\theta_2 (t, z) = 0$, with the first phase representing 
a freely propagating acoustic wave along the $z-$axis, and 
the second phase is a trivial (constant) solution that shows 
no acoustic wave; these results are consistent with a more 
general (3-dimensional) solution given by Eqs (\ref{eq29}) 
and (\ref{eq30}).  The obtained results demonstrate that 
the propagation of very high frequency acoustic waves is 
not affected by stratification of the isothermal atmosphere.

The effects of medium stratification on the acoustic wave 
propagation become important when $\omega \gtrsim 
\Omega_{ac}$; in this case, the wave phase is given by 
Eq. (\ref{eq35}) and both solutions contribute to $\phi 
(z,t)$ given by Eq. (\ref{eq36}).  The most interesting 
case is when $\omega = \Omega_{ac}$, which gives 
$\theta_{1,2} (t, z) = \pm (2z/v \pm \vert \hat{z} \cdot 
\hat {v} \vert t) \Omega_{ac}$, showing that
propagating acoustic waves cease to exist as they are 
replaced by oscillations of the atmosphere with its 
natural frequency $\Omega_{ac}$.  The existence of 
oscillations in planetary, solar and stellar atmospheres 
is well known [47-49].  The origin of solar 5-min oscillations 
is attributed to the acoustic waves trapped in the solar interior 
[48]; however, the 3-min oscillations of the solar atmosphere
are driven by the propagating acoustic waves [50].  The results 
presented in this paper demonstrate that the propagation of 
acoustic waves is terminated when $\omega = \Omega_{ac}$, 
and that the solar atmosphere begins to oscillate with its natural
frequency $\Omega_{ac}$, which is also the cutoff frequency 
for acoustic waves, as it was first shown by Lamb [39-41].  

Having demonstrated that acoustic wave propagation 
is terminated in the limit when $\omega \rightarrow 
\Omega_{ac}$, this means that $\Omega_{ac}$ is the 
Lamb (acoustic) cutoff frequency.  It must be now verified 
that the wave phases given by Eq. (\ref{eq35}) become 
imaginary for any $\omega < \Omega_{ac}$, that is, 
the solutions $\phi_{1,2} (t, z)$ are exponentially 
decaying and the waves are evanescent.  If $\omega 
\lesssim \Omega_{ac}$, the wave phases are imaginary 
and given by  
\begin{equation}
\theta_{1,2} (t, z) = i \frac{c_{s}}{v} \left [ 
\sqrt{\Omega_{ac}^2 - \omega^2} \pm \sqrt{\left ( 1 - 
\frac{v^2} {c_{s}^2} \right ) \Omega_{ac}^2 - 
\omega^2} \right ] \left ( \frac{2z}{v} \pm  
\vert \hat{z} \cdot \hat {v} \vert  t \right )
\label{eq37}
\end{equation}
In general, the term $[(1 - v^2 / 4 c_s^2) \Omega_{ac}^2 
- \omega^2] > 0$, but it may also become negative if $v > 
2 c_s$, which means that if the second term of these phases 
becomes imaginary, then this term would give oscillatory 
solutions.  However, this does not affect the general solution 
as the exponential decay caused by the first term takes over 
and makes the entire solution evanescent.  Similarly, when 
$\omega << \Omega_{ac}$, the wave phases become 
\begin{equation}
\theta_{1,2} (t, z) = i \frac{c_{s}}{v} \left [\Omega_{ac}
\pm \sqrt{\left ( 1 - \frac{v^2} {c_{s}^2} \right )}
\Omega_{ac} \right ] \left (\frac{2 z}{v} \pm 
\vert \hat{z} \cdot \hat {v} \vert  t \right )\ ,
\label{eq38}
\end{equation}
showing that the solutions are exponentially decaying. 
Based on the above discussion, the obtained results are 
valid in both cases when $v \leq 2 c_s$ as well as when 
$v > 2 c_s$.  Thus, acoustic waves of all frequencies 
lower than $\Omega_{ac}$ are always evanescent.  

The presented results show that the FWE for acoustic waves
can be derived from the hydrodynamic equations after using 
the eigenvalue equation given by Eq. (\ref{eq1b}).  As a result, 
the FWE directly displays the characteristic atmospheric frequency 
$\Omega_{ac}$ similar as the AWE does.  By solving the FWE 
for acoustic waves, it is demonstrated that $\Omega_{ac}$ is 
the Lamb (acoustic) cutoff frequency that uniquely determines 
the conditions for the acoustic wave propagation, which is 
consistent with the original results presented by Lamb in 
1910 [39].  However, there are main differences between 
the results presented in this paper and those obtained by 
Lamb [39-41], namely, Lamb's results are valid in only one 
stationary inertial frame $S$, which is selected to describe 
the waves, while the presented results are the same for all 
inertial observers in the Galilean space and time.  In other 
words, for all Galilean observers, the waves have the same 
wave speed, frequency and wavenumber, and their propagation 
conditions remain also frame-independent, which shows that 
the newly formulated theory of acoustic waves based on the 
FWE is fundamental.

In realistic physical situations when the wave speed is not 
constant, and wave damping and nonlinearities may be present, 
the FWE may lose its status of being fundamental.  Nevertheless, 
it will still remain another wave equation, which may be applicable 
to some physical situations involving classical waves as the 
Schr\"odinger equation has been used [19-23]. 

\section{Conclusions}

A method based on the irreps of the extended Galilean group
is used to derive infinite sets of symmetric and asymmetric 
second-order PDEs with constant coefficients of arbitrary real 
values.  The obtained results demonstrate that among these 
equations only one asymmetric equation is a new fundamental 
wave equation, which gives the most complete description 
of propagating waves as it accounts for the Doppler effect, 
forward and backward waves, and makes the wave speed 
to be the same in all inertial frames.  Thus, the main 
result of this paper is that classical waves propagating with 
speeds $c_w << c$ may 'mimic' the behavior of electromagnetic 
waves when they are described by the FWE.  It is suggested 
that these waves be called the {\it basic classical waves} in 
Galilean Relativity because they 'mimic' the behavior of EM 
waves in Special Theory of Relativity.

Contrary to the standard wave equation and the Schr\"odinger 
equation for classical waves, which are second-order PDEs, the 
new fundamental asymmetric wave equation discovered in this 
paper is an ODE. The conversion from the PDE to ODE was 
achieved by using wave phases as the independent wave 
variables that depend on both time and space.  An interesting 
result of this paper is that only the new asymmetric equation 
can be converted into the fundamental wave equation, and 
that its form resembles the law of inertia. The mathematical 
forms of both equations are similar; however, the new 
fundamental asymmetric wave equation has one extra term 
that allows for periodic solutions.  This may suggest that 
the derived fundamental wave equation plays the same 
role for classical waves in theories of waves as the law 
of inertia plays for classical particles in CM.

The fundamental wave equation is applied to the propagation of 
acoustic waves in an isothermal atmosphere.  The analysis shows 
that the wave propagation conditions are uniquely determined 
by the existence of the atmospheric natural frequency, which is 
identified with the acoustic cutoff frequency originally introduced 
by Lamb [39].  However, while Lamb's wave description and its 
cutoff frequency are frame-dependent, the wave description given 
by the new fundamental wave equation (Eq. \ref{eq32}) and its 
acoustic cutoff remain the same for all Galilean observers in their 
inertial frames.  The presented theory of waves based on the 
fundamental wave equation also predicts the existence of 
atmospheric oscillations with the natural atmospheric frequency 
that are driven by the process of the propagating waves becoming 
evanescent when their frequencies become equal to the Lamb 
frequency.   

\bigskip\noindent
{\bf Acknowledgment:} 
I appreciate very much valuable comments and several 
stimulating questions asked by two anonymous referees, 
which allowed me to significantly improve the original 
version of this paper.  The author also thanks Dora 
Musielak for comments and suggestions on the earlier 
version of this manuscript.  This work was partially 
supported by Alexander von Humboldt Foundation.\\

\appendix

\section{Derivation of the eigenvalue equations}

Let us consider a set of $N$ functions that forms a basis of an N-dimensional representation 
given by a set of $N \times N$ matrices A for each irrep, and for each element of the group  
\begin{equation}
\hat \alpha f_{l}^{(k)}\ =\ \sum_m A_{ml}(\hat \alpha) 
f_{m}^{(k)}\ ,
\label{eqA1}
\end{equation}
where $\alpha$ is one of the elements of the group, $k$ labels the irreps and $l$ is one of 
the members of the set of N functions satisfying Eq. (A1). In addition, the sum on $m$ is 
over the N members of the set, and the matrices $A$ are unitary. 

Writing Eq. (\ref{eqA1}) for space translations $\mathbf {a}$, the result is   
\begin{equation}
\hat T_{\mathbf a} \psi(t, \mathbf x)\ \equiv\ \psi(t, \mathbf x + \mathbf a)\ 
=\ e^{i \mathbf {k} \cdot \mathbf {a}} \psi (t, \mathbf x)\ .
\label{eqA2}
\end{equation}
Making the Taylor series expansion of $\phi (\mathbf {r} + \mathbf {a})$, one gets 
\begin{equation}
\phi (t, \mathbf {x} + \mathbf {a}) = \exp [ i (- i \mathbf {a} \cdot \nabla) ] 
\phi (t, \mathbf x)\ .
\label{eqA3}
\end{equation}
Comparing Eq. (\ref{eqA3}) to Eq. (\ref{eqA2}), the following eigenvalue 
equation is obtained 
\begin{equation}
- i \nabla \phi (t, \mathbf {x}) = \mathbf {k}\ \phi (t, \mathbf {x})\ ,
\label{eqA4}
\end{equation}  
which is the eigenvalue equation given by Eq. (\ref{eq1b}).

For the time translation $t_0$, one obtains 
\begin{equation}
\hat T_{t_0} \psi(t, \mathbf x)\ \equiv\ \psi(t + t_0, \mathbf x)\ =\
e^{- i \omega t_0} \psi (t, \mathbf x)\ .
\label{eqA5}
\end{equation}
Comparison of this equation to the Taylor series expansion 
\begin{equation}
\phi (t + t_0, \mathbf {x}) = \exp [ i ( - i t_0\ \partial / \partial t) ] 
\phi (t, \mathbf x)\ .
\label{eqA6}
\end{equation}
gives
\begin{equation}
i {{\partial} \over {\partial t}} \phi (t, \mathbf {x}) = \omega\ 
\phi (t, \mathbf {x})\ ,
\label{eqA7}
\end{equation}  
which is the eingenvalue equation given by Eq. (\ref{eq1a}).

The derived eigenvalue equations represent the necessary conditions 
that $\phi (t, \mathbf {x})$ transforms as one of the irreps of $T(3+1)$
[17]. The above results also show that the irreps of the group $T(3+1)$ 
are labeled by the real vector $\mathbf {k}$ and the real scalar 
$\omega$, and there are no other restrictions on these quantities. 
It must also be mentioned that these labels are preserved in the 
irreps of the entire $\mathcal {G}_e$ because $T(3+1)$ is its 
invariant subgroup [10].

\end{document}